\documentstyle[12pt]{article}

\begin{document}
\newcommand{\nd}[1]{/\hspace{-0.5em} #1}
\begin{titlepage}
\begin{flushright}
hep-th/9607066 \\
SWAT/114 \\
DTP/96/58 \\
\end{flushright}
\begin{centering}
\vspace{.2in}
{\Large {\bf A Two-Instanton Test of the 
Exact Solution\\ of $N=2$ Supersymmetric QCD}}
\vspace{.5in}

{\bf Nicholas Dorey }\\
\vspace{.1in}
Physics Department, University College of Swansea \\
Swansea, SA2 8PP, UK\\
\vspace{.5in}
{\bf Valentin V. Khoze} \\
\vspace{.1in}
Physics Department, Centre for Particle Theory  \\
 University of Durham, Durham DH1 3LE, UK \\
\vspace{.5in}
{\bf Michael P. Mattis} \\
\vspace{.1in}
Theoretical Division,  Los Alamos National Laboratory \\
Los Alamos, NM 87545, USA\\
\vspace{.6in}
%
The exact solution of $N=2$ supersymmetric QCD found by Seiberg and
Witten includes a prediction for all multi-instanton contributions to
the low-energy dynamics. In the case where the matter hypermultiplets
are massless, the leading non-perturbative contribution at weak
coupling comes from the two-instanton sector. We calculate this 
contribution from first principles using
the exact two-instanton solution of the self-dual Yang-Mills equation 
found by Atiyah, Drinfeld, Hitchin and Manin. 
We find exact agreement with the predictions of Seiberg and
Witten for all numbers of flavours. We also confirm their predictions 
for the case of massive hypermultiplets. 
\end{centering}

\end{titlepage}
\section*{}
\paragraph{}
In recent years, the idea of electric-magnetic duality has lead to a
dramatic improvement in our understanding of gauge theories with
extended supersymmetry. Due to the work of Seiberg and Witten
\cite{SW1,SW2}, and 
subsequent generalizations, exact results are available both for the
spectrum of BPS-saturated states and for the  
low-energy effective action. 
A common feature of these results is that they have highly non-trivial
consequences at weak coupling which can be compared with the results
of semiclassical calculations. In the case of
$N=2$ supersymmetric Yang-Mills (SYM) theory with gauge group $SU(2)$,
the exact low-energy effective action obtained in 
\cite{SW1} contains predictions for all
multi-instanton contributions to low-energy Green functions. 
These predictions have been checked by explicit calculation in the 
 one-\cite{FP} and two-instanton \cite{MO} sectors. 
In this letter we extend these calculations to the $N=2$ theory
coupled to $N_{F}$ 
hypermultiplets of matter in the fundamental representation. 
Our results provide a semiclassical test 
of the exact solution of this model at low-energy proposed in \cite{SW2}. 
The analysis relies heavily on the results for SYM theory (ie
$N_{F}=0$) derived in \cite{MO} and we will concentrate primarily on
the new features which appear in the presence of matter
hypermultiplets. A more complete account of these calculations will
appear elsewhere \cite{MOII}.  
\paragraph{}
We begin by restricting our attention to the case of massless hypermultiplets. 
In this case, a discrete 
symmetry ensures that only even numbers of instantons contribute \cite{SW2}. 
For $N_{F}<4$, where the theory is asymptotically free, 
the exact low-energy effective action, including all
terms with at most two derivatives or four fermions, can be derived from a
holomorphic prepotential which has the following expansion at weak
coupling \cite{SW2,IT}, 
\begin{equation}
{\cal F}^{(N_{F})}(\Psi)=i\frac{(4-N_{F})}{8\pi}\Psi^{2}
\log{\left(\frac{C\,\Psi^{2}}{\Lambda^{2}_{N_{F}}}\right)}
-\frac{i}{\pi}\sum_{k=1}^{\infty}{\cal F}^{(N_{F})}_{2k}\left(
\frac{\Lambda_{N_{F}}}{\Psi}\right)^{2k(4-N_{F})}\Psi^{2}
\label{pre}
\end{equation}
where $\Psi$ is an $N=2$ chiral superfield containing the massless photon and
its superpartners,
$\Lambda_{N_{F}}$ is the dynamically generated
scale of the theory,\footnote{The numerical values of the constants
${\cal F}^{(N_{F})}_{2k}$ depend on the definition of the scale 
$\Lambda_{N_{F}}$. In this paper, as in \cite{MO}, we adopt the 
$\Lambda$-parameter of the Pauli-Villars regularization scheme which
is appropriate for instanton calculations. This is related to the
$\Lambda$-parameter of \cite{SW1,SW2} as 
$4\Lambda^{4-N_{F}}_{N_{F}}=(\Lambda^{SW}_{N_{F}})^{4-N_{F}}$ \cite{FP}.} 
and $C$ is a numerical constant.
The logarithmic term comes from one-loop
perturbation theory while the $k$-th term in the infinite series
corresponds to the leading semiclassical contribution of $2k$
instantons. In the following we will only consider the
first term in this series: the two-instanton contribution. Including
the result for $N_{F}=0$,\footnote{Note that, for $N_{F}=0$, the
prepotential has contributions from all numbers of instantons.}  
a straightforward Taylor expansion of the exact solution specified in 
\cite{SW2}, yields the numerical values 
${\cal F}^{(0)}_{2}=5/2^{4}$, ${\cal F}^{(1)}_{2}=-3/2^{5}$
and ${\cal F}^{(2)}_{2}=1/2^{6}$. 
The coefficient ${\cal F}^{(3)}_{2}$ 
corresponds to an additive constant in the prepotential which does not
contribute to the low-energy effective Lagrangian and is irrelevant
for our purposes. Following \cite{S,FP,MO}, we will check these values by
comparing the corresponding contributions 
to the four anti-fermion Green function  
$G^{(4)}_{N_{F}}(x_{1},x_{2},x_{3},x_{4})=\langle
\bar{\lambda}(x_{1})\bar{\lambda}(x_{2})\bar{\psi}(x_{3})
\bar{\psi}(x_{4})\rangle$ with the result of an explicit two-instanton
calculation. 
\paragraph{}
We also consider the case $N_{F}=4$, where 
the perturbative 
$\beta$-function vanishes. In this case the theory is believed to have
an exact electric-magnetic duality which requires that all
non-perturbative corrections to the low-energy effective action should also
vanish \cite{SW2}. In the following, 
we explicitly check the vanishing of the two-instanton contribution to
$G^{(4)}_{4}$. Finally we will
briefly describe how these results are modified by the introduction of
masses for the hypermultiplets.    
\paragraph{}
In the absence of matter, the particle content of the $N=2$ theory is
conveniently described in terms of two $N=1$ superfields which
transform in the adjoint representation of the gauge group. The gauge
superfield $V$ contains the Yang-Mills gauge field $v_{m}$ and the
gaugino $\lambda$, while the complex chiral multiplet $\Phi$
contains the adjoint Higgs $A$ and the Higgsino $\psi$. The theory has
flat directions along which the scalar field $A$ acquires a vacuum
expectation value (VEV), $A_{\rm cl}$, breaking the gauge group down to an 
abelian subgroup: $SU(2)\rightarrow U(1)$. Without loss of generality 
we may choose $A_{\rm cl}={\rm v}\tau^{3}/2$ where ${\rm v}$ is an 
arbitrary complex number which parametrizes the moduli space of 
gauge-inequivalent vacua. In addition we
introduce $N_{F}$ matter hypermultiplets which transform in the fundamental
representation of $SU(2)$. Each $N=2$ hypermultiplet corresponds to a pair of
$N=1$ chiral multiplets, $Q_{i}$ and $\tilde{Q}_{i}$ where 
$i=1,2\ldots N_{F}$, which contain scalar quarks (squarks) $q_{i}$ and
$\tilde{q}_{i}$ respectively and fermionic partners 
$\chi_{i}$ and $\tilde{\chi}_{i}$. 
Importantly we are restricting our attention to the Coulomb branch of
the theory where the squarks do not acquire a VEV.  
In the $N=1$ language, the matter fields couple to the gauge multiplet
via a superpotential, 
\begin{equation}
W=\sum_{i=1}^{N_{F}} \sqrt{2}\tilde{Q}_{i}\Phi Q_{i} +
m_{i}\tilde{Q}_{i}Q_{i}
\label{superp}
\end{equation}
suppressing colour indices. 
We have also included masses $m_{i}$ for the
hypermultiplets. Until further notice we set $m_{i}=0$.
\paragraph{}
As in \cite{MO}, we begin by constructing the short-distance
superinstanton.
The short-distance behaviour of the gauge field 
is determined by solving the self-dual Yang-Mills equation, while the 
fermion fields obey the covariant Dirac equation in the self-dual
background. We must also solve the Euler-Lagrange equations for the scalar
fields, subject to vacuum boundary conditions. 
In contrast, the large-distance behaviour of the massless fields is
obtained by solving the linearized equations appropriate to 
the low-energy abelian theory. These equations yield power-law
behaviour with arbitrary coefficients which must be fixed by matching
onto the short-distance solutions \cite{ADS}. 
As usual the resulting field configurations are not solutions of the 
exact coupled 
equations of motion but should be thought of as approximate solutions
corresponding to quasi-flat directions in configuration space \cite{AF}.  
\paragraph{}
The general 
solution of the self-dual Yang-Mills equation was found by Atiyah, 
Drinfeld, Hitchin and Manin (ADHM) \cite{ADHM} and is conveniently given 
in terms matrices with quaternionic entries
\footnote{Quaternions are represented as 
$2\times 2$  matrices; e.g., $B^{\alpha\dot{\alpha}}=
B_{m}\sigma^{\alpha\dot{\alpha}}_{m}$, $m=0,\ldots,3$ where 
$\sigma^{\alpha\dot{\alpha}}_{m}$ are the $\sigma$-matrices of  
Wess and Bagger \cite{WB}. Wherever there is no ambiguity we suppress
these indices. For a full list of our 
conventions see Sec.~6 of \cite{MO}.}. Specifically, the
solution for the gauge field $v_{m}$ with topological charge $n$ 
is given by, 
\begin{eqnarray}
v_{m}=\bar{U}_{\lambda}\partial_{m}U_{\lambda} &\qquad{} \qquad{}  &
\bar{U}_{\lambda}U_{\lambda}=I   
\label{ADHM1}
\end{eqnarray}
where $I$ is the unit quaternion and $U$ 
is an $x$-dependent $(n+1)$-component column vector with 
quaternionic entries which obeys,    
\begin{eqnarray} 
\bar{\Delta}_{k\lambda}U_{\lambda}=0 & & \Delta_{\lambda k}=
a_{\lambda k}+b_{\lambda k}x 
\label{ADHM2}
\end{eqnarray}
where $a$ and $b$ are $(n+1)\times n$ matrices of 
quaternionic constants. These constant matrices are further restricted by
a nonlinear constraint which can be written as
\begin{equation}
\bar{\Delta}_{k\lambda}\Delta_{\lambda l}\ =\ \big(f^{-1}\big)_{kl}\,I
\label{ADHM3} 
\end{equation}
This equation constrains $a$ and $b$ by
demanding that each entry in the matrix product on the LHS be 
proportional to the unit quaternion for all $x$. In general, after
implementing the constraint and eliminating all redundant degrees of
freedom the resulting solution depends on $8n$ independent
parameters. In an asymptotic limit, these parameters can be identified
as the positions, scale sizes and global gauge orientations of $n$ 
well-separated BPST instantons.  
\paragraph{}
Specializing to the case of topological charge $n=2$, the canonical 
form for the ADHM solution is specified by the following
choices for $a$ and $b$;
\begin{eqnarray}
a=\left(\begin{array}{cc} w_{1} & w_{2} \\ x_{0}+a_{3} & a_{1} \\ a_{1}
& x_{0}-a_{3} \end{array}\right) & \qquad{} \qquad{} & 
b=\left(\begin{array}{cc} 0 & 0 \\ I & 0 \\ 0
& I \end{array}\right)
\label{ab}
\end{eqnarray}
In general there are an infinite number of ways of solving the constraint
(\ref{ADHM3}).  A particularly convenient choice is to eliminate the
off-diagonal element $a_{1}$ as \cite{CWS}; 
\begin{equation}
a_{1}=\frac{1}{4|a_{3}|^{2}}a_{3}(\bar{w}_{2}w_{1}-\bar{w}_{1}w_{2})
\label{a1} 
\end{equation}
In the limit $|a_{3}|\rightarrow\infty$, the corresponding field
configuration resembles two well-separated instantons centered at 
$-(x_{0}+a_{3})$ and $-(x_{0}-a_{3})$, with scale-sizes $|w_{1}|$ and
$|w_{2}|$ respectively.     
\paragraph{}
At short distance, 
the adjoint fermions $\lambda$ and $\psi$ take their zero modes values
in the self-dual background. The complete set of $4n$ linearly
independent solutions of the adjoint Dirac equation in the general
ADHM background was found by Corrigan et al
\cite{COR1}. The general solution for each species of adjoint fermion
depends on an $(n+1)\times n$ matrix 
with constant Grassmann spinor entries. Following Sec.~7.1 of \cite{MO}, 
we denote these matrices ${\cal M}_{\gamma}$ and ${\cal N}_{\gamma}$
for the gaugino $\lambda$ and the Higgsino $\psi$, respectively.   
The matrices ${\cal M}_{\gamma}$ and ${\cal N}_{\gamma}$
satisfy the following linear constraints, 
\begin{eqnarray}
\bar{a}^{\dot{\alpha}\gamma}{\cal M}_{\gamma}=-{\cal M}^{\gamma T}
a_{\gamma}^{\dot{\alpha}} & \qquad{} & 
\bar{a}^{\dot{\alpha}\gamma}{\cal N}_{\gamma}=-{\cal N}^{\gamma T}
a_{\gamma}^{\dot{\alpha}}
 \label{fcon1} \\ 
\bar{b}_{\alpha}^{\ \gamma}{\cal M}_{\gamma}={\cal M}^{\gamma T}
b_{\gamma\alpha} & & \bar{b}_{\alpha}^{\ \gamma}{\cal
N}_{\gamma}={\cal N}^{\gamma T}
b_{\gamma\alpha}
\label{fcon2}
\end{eqnarray}
where we have suppressed the ADHM indices.  
After imposing the constraints we find a total of $8n$ independent
Grassmann parameters. These parameters are fermionic collective coordinates   
which must be integrated over. In order to find the correct measure
for this integration, it is necessary to evaluate the $8n\times 8n$
normalization matrix for the adjoint zero modes. 
This is accomplished in Appendix 
B of \cite{MO}. 
\paragraph{}
Specializing to the case $n=2$, the
matrices ${\cal M}_{\gamma}$ and ${\cal N}_{\gamma}$ are parametrized as,  
\begin{eqnarray}
{\cal M}_{\gamma}=\left(\begin{array}{cc} \mu_{1\gamma} & \mu_{2\gamma} \\ 
4\xi_{\gamma}+{\cal M}_{3\gamma} & {\cal M}_{1\gamma} \\ {\cal M}_{1\gamma}
& 4\xi_\gamma-{\cal M}_{3\gamma} \end{array}\right) & & 
{\cal N}_{\gamma}=\left(\begin{array}{cc} \nu_{1\gamma} & \nu_{2\gamma} \\ 
4\xi'_{\gamma}+{\cal N}_{3\gamma} & {\cal N}_{1\gamma} \\ {\cal N}_{1\gamma}
& 4\xi'_\gamma-{\cal N}_{3\gamma} \end{array}\right) \nonumber
\label{mn}
\end{eqnarray}
This form already satisfies the  
constraint (\ref{fcon2}). Just like its bosonic counterpart in 
(\ref{ADHM3}), the remaining fermionic constraint (\ref{fcon1}) is solved by
eliminating the off-diagonal elements, ${\cal M}_{1\gamma}$ and 
${\cal N}_{1\gamma}$; 
\begin{eqnarray}
{\cal M}_{1} &= & \frac{1}{2|a_{3}|^{2}}a_{3}(2\bar{a}_{1}
{\cal M}_{3}+
\bar{w}_{2}\mu_{1}-\bar{w}_{1}\mu_{2}) \nonumber \\
{\cal N}_{1} &= & \frac{1}{2|a_{3}|^{2}}a_{3}(2\bar{a}_{1}
{\cal N}_{3}+
\bar{w}_{2}\nu_{1}-\bar{w}_{1}\nu_{2})
\label{m1n1} 
\end{eqnarray}
The fermionic zero modes parametrized by $\xi_{\alpha}$ and
$\xi'_{\alpha}$ correspond to the four supersymmetry 
generators which act non-trivially on the self-dual gauge field.
\paragraph{}
In addition to the adjoint fermions which belong to the gauge
multiplet, we must also consider the zero modes of the fundamental
fermions $\chi_{i}$ and $\tilde{\chi}_{i}$, for $i=1,\ldots, N_{F}$, 
in the ADHM background. 
For arbitrary instanton number, these modes are given as \cite{COR2,O1},    
\begin{eqnarray}
\left(\chi^{\alpha}_{i}\right)^{\dot{\beta}} & = & 
\bar{U}^{\dot{\beta}\alpha}_{\lambda}b_{\lambda k}f_{kl}{\cal K}_{li}
\nonumber \\
\left(\tilde{\chi}^{\alpha}_{i}\right)^{\dot{\beta}} & = & 
\bar{U}^{\dot{\beta}\alpha}_{\lambda}b_{\lambda
k}f_{kl}\tilde{\cal K}_{li}
\label{fund}
\end{eqnarray}
with $\alpha$ a Weyl and $\dot\beta$ an $SU(2)$ colour index.
The normalization matrix of these modes is diagonal in ADHM
indices \cite{O1}, 
\begin{equation}
\int\, d^{4}x (\chi^{\alpha}_{i})^{\dot{\beta}}
(\tilde{\chi}_{\alpha j})_{\dot{\beta}}= 
\pi^{2}{\cal K}_{li}\tilde{\cal K}_{lj} 
\label{orth}
\end{equation}
In the two instanton case we have a total of $4N_{F}$ linearly
independent zero modes, corresponding to the Grassmann parameters
${\cal K}_{li}$ and $\tilde{\cal K}_{li}$ with $l=1,2$. 
For later convenience we define the following, frequently occurring, 
combinations of the collective coordinates introduced above; 
\begin{eqnarray}
L &= & |w_{1}|^{2}+|w_{2}|^{2} \nonumber \\
H &= & |w_{1}|^{2}+|w_{2}|^{2} + 4|a_{1}|^{2}+4|a_{3}|^{2} \nonumber \\
Y & = & \mu_{1}\nu_{2}-\nu_{1}\mu_{2}+2{\cal M}_{3}{\cal N}_{1}
-2{\cal N}_{3}{\cal M}_{1} \nonumber \\ 
Z & = & \sum_{i=1}^{N_{F}}
{\cal K}_{ki}\epsilon_{kl}\tilde{\cal K}_{li} 
\label{useful}
\end{eqnarray}
\paragraph{}
Following \cite{MO} we next 
solve the equation of motion for the adjoint scalar fields
in the presence of the fermionic sources described above
\footnote{These equations follow directly from the Lagrangian (2.3) of
\cite{MO} with couplings to matter implied by the
superpotential (\ref{superp}).};
\begin{eqnarray}
({\cal D}^{2}A)^{\dot{\gamma}}_{\,\,\,\dot{\alpha}} & = & i\sqrt{2}g\,
\big(
 \lambda^{\dot{\gamma}}_{\,\,\,\dot{\beta}}
\psi^{\dot{\beta}}_{\,\,\,\dot{\alpha}}
-
\psi^{\dot{\gamma}}_{\,\,\,\dot{\beta}}
\lambda^{\dot{\beta}}_{\,\,\,\dot{\alpha}}\big)
 \\
({\cal D}^{2}A^{\dagger})^{\dot{\gamma}}_{\,\,\,\dot{\alpha}} & = & 
\frac{g}{2\sqrt{2}}\,\sum_{i=1}^{N_{F}}(\chi_i^{\dot{\gamma}}
\tilde{\chi}^{}_{i\dot{\alpha}}+
\tilde{\chi}_i^{\dot{\gamma}}\chi^{}_{i\dot{\alpha}}
)
\label{lap2} 
\end{eqnarray}
displaying colour and flavour but suppressing Weyl indices. 
Note that, in the path integral, the fermion bilinear contributions to 
$A$ and to $A^{\dagger}$ are completely independent.  
The above equations must be solved subject to the vacuum boundary conditions
$A(x)\rightarrow A_{\rm cl}={\rm v}\tau_{3}/2$ as 
$|x|\rightarrow\infty$. In \cite{MO}, we
described the method for solving these equations in detail for the
case $N_{F}=0$. The only difference in the present case is the
presence of a source term on the RHS of (\ref{lap2}) 
involving the fundamental matter fermions. This term can be included
simply by an appropriate redefinition of the constant parameters
appearing in the solution for $N_{F}=0$. Explicitly we find, 
\begin{eqnarray}
iA &= &\bar{U}_{\lambda}({\cal A}+
{\cal B})_{\lambda\lambda'}U_{\lambda '} \\
iA^{\dagger} &= &
\bar{U}_{\lambda}(\bar{\cal A}+
\bar{\cal C}_{\scriptscriptstyle\rm })_{\lambda\lambda'}
U_{\lambda'}
\end{eqnarray}
Here $\bar{\cal A}({\rm v})=-{\cal
A}(\bar{\rm v})$ and 
\begin{eqnarray}
{\cal A}_{\lambda\lambda'} & = & i\delta_{\lambda 0}\delta_{\lambda'
0} A_{\rm cl} + {\rm k}\,
b_{\lambda k}\epsilon_{kl}\bar{b}_{\l\lambda'}
\\
{\cal B}_{\lambda \lambda '} & = & \frac{1}{2\sqrt{2}}({\cal N}f
{\cal M}^{T}-{\cal M}f{\cal N}^{T})_{\lambda\lambda '} + 
\kappa\,b_{\lambda k}\epsilon_{kl}\bar{b}_{\l\lambda'} \\
\bar{\cal C}_{\lambda\lambda'}
 &= & \bar{\rho}\,
b_{\lambda k}\epsilon_{kl}
\bar{b}_{\l\lambda'}  
\label{adef}
\end{eqnarray}
where 
\begin{equation}
{\rm k}={\omega\over H}  \qquad
\kappa=-\frac{g}{2\sqrt{2}} \frac{Y}{H}   \qquad
\bar{\rho}= \frac{ig}{8\sqrt{2}}\frac{Z}{H}
\label{defk}
\end{equation}
with
$\omega=\bar{w}_{2}A_{\rm cl}w_{1}-\bar{w}_{1}A_{\rm cl}w_{2}$.
\paragraph{}
As mentioned above, these field configurations are not exact solutions
to the coupled equations of motion. It follows that the Euclidean action
evaluated on the configurations found above will depend explicitly on the
collective coordinates. This dependence has two important effects. 
First it provides an effective IR cut-off on the integral over
instanton scale sizes. Second, the dependence on fermionic collective
coordinates lifts a subset of the fermion zero modes. For a SUSY gauge 
theory with an adjoint scalar, it turns out that all fermion zero
modes are lifted except those which are protected by supersymmetry \cite{MO}. 
In the present case, as for $N_{F}=0$, we will find that the only 
unlifted modes are the four modes parametrized by 
$\xi_{\alpha}$ and $\xi'_{\alpha}$ identified above.  
\paragraph{}
In \cite{MO}, we obtained an exact expression for the super-instanton
action for the case $N_{F}=0$. In that case, it was possible, using
integration by parts and the equations of motion, to express the total action 
as a surface term. The resulting action 
was then determined soley by the $1/x^{3}$ term in the expansion of 
adjoint scalar field at large distance. The final answer (Eqn (8.7) of
\cite{MO}) contained a bosonic
part and a part bilinear in the adjoint Grassmann parameters 
${\cal M}_{\gamma}$ and ${\cal N}_{\gamma}$. 
In the present case, the same method can be
used to calculate the bosonic and fermion bilinear terms in the
action. However there is also a new feature for $N_{F}>0$: a term
quadrilinear in the Grassmann parameters ${\cal M}_{\gamma}$, 
${\cal N}_{\gamma}$, ${\cal K}_{i}$ and $\tilde{\cal K}_{i}$, which
cannot be written as a surface term in any obvious way. Instead this term
is uniquely determined by the requirement that the total action should
be a supersymmetric invariant \cite{MOII}. 
Including this term, the super-instanton action 
is given as $S_{N_{F}}=16\pi^{2}/g^{2}+\tilde{S}_{N_{F}}$ where 
\begin{equation}
\tilde{S}_{N_{F}}=4\pi^{2}L|{\rm v}|^{2}-
16\pi^{2}(\bar{\omega}+\lambda Z)({\rm k}+\kappa)
+4\sqrt{2}\pi^{2}ig\nu_{k}\bar{A}_{\rm cl}\mu_{k}
\label{action}
\end{equation}
with $\lambda=ig/8\sqrt{2}$. The only difference from the
$N_{F}=0$ expression is the dependence on the fundamental parameters
${\cal K}_{i}$ and $\tilde{\cal K}_{i}$ introduced by the replacement
\footnote{In contrast it is easy to show that there is no dependence on the
parameters ${\cal K}_{i}$ and $\tilde{\cal K}_{i}$ in the action of a 
single superinstanton. It follows that the Grassmann integrals over
these parameters in the one-instanton measure are not
saturated and the corresponding contribution to $G_{N_{F}}^{(4)}$
vanishes as expected from (\ref{pre}).} 
$\bar{\omega}\rightarrow \bar{\omega}+\lambda Z$.
\paragraph{}
At large-distances, $|x|\gg 1/g|{\rm v}|$, the massless components of the 
anti-fermion have a power-law decay with a coefficient which can be
related to the collective coordinate action (\ref{action}) \cite{MO}
\footnote{In particular, see the more
recent, slightly expanded, version of \cite{MO} where all expressions
are given in the general case of complex ${\rm v}$.}:    
\begin{eqnarray}
\bar{\lambda}^{\scriptscriptstyle\rm LD}_{\dot{\alpha}}(x) & = & 
i\sqrt{2}\,\frac{\partial S_{N_{F}}}
{\partial {\rm v}}\xi'^{\alpha} {\rm S}_{\alpha\dot{\alpha}}(x,x_{0}) 
\nonumber
\\ 
\bar{\psi}^{\scriptscriptstyle\rm LD}_{\dot{\alpha}}(x) 
& = & i\sqrt{2}\,\frac{\partial S_{N_{F}}}
{\partial {\rm v}}\xi^{\alpha} {\rm S}_{\alpha\dot{\alpha}}(x,x_{0})
\label{larged}
\end{eqnarray}
where ${\rm S}_{\alpha\dot{\alpha}}$ is the free propagator for a massless
Weyl fermion. Here we are treating 
${\rm v}$ and $\bar{\rm v}$ as independent 
variables and using the fact that the action (\ref{action}) is linear
in each of these variables. 
Note that these fields now have a contribution trilinear in the 
Grassmann collective coordinates.  
\paragraph{}
The final ingredient in our calculation is the collective coordinate 
measure \cite{O2}. An explicit expression in the case $N_{F}=0$ was given in 
Sec.~8.1 of \cite{MO}. 
Apart from the modification of the two-instanton action 
described above, the only new feature here is the integration over the
parameters ${\cal K}_{li}$ and $\tilde{\cal K}_{li}$. The accompanying
Jacobian follows immediately from the normalization relation
(\ref{orth}). For $N_{F}$ flavours we have, 
\begin{eqnarray}
\int d\mu^{(N_{F})}_{2} & =& \frac{2^{10}
\Lambda^{2(4-N_{F})}_{N_{F}}}{\pi^{4(2+N_{F})}{\cal S}_{2}}
\int\, d^{4}x_{0}d^{4}a_{3}d^{4}w_{1}d^{4}w_{2} \, \times \,
 d^{2}\xi d^{2}{\cal M}_{3}d^{2}\mu_{1}d^{2}\mu_{2} \nonumber \\ 
 & \, \times \, &
d^{2}\xi'd^{2}{\cal N}_{3}d^{2}\nu_{1}d^{2}\nu_{2} \, \times \, 
\prod_{i=1}^{N_{F}} 
d{\cal K}_{1i}d{ \cal K}_{2i}d\tilde{\cal K}_{1i}d\tilde{\cal K}_{2i}
\nonumber \\
& \,\times\, &  \exp\left(-\tilde{S}_{N_{F}}\right)\,
\frac{\left| |a_{3}|^{2}-|a_{1}|^{2}\right|}{H}  \nonumber
\label{measure}
\end{eqnarray}
where ${\cal S}_{2}=16$ is a symmetry factor associated with a
discrete redundancy in the chosen parametrization of the two instanton
solution \cite{MO,O2}. 
\paragraph{}
Putting the pieces together,  
we find for the four-point function; 
\begin{eqnarray}
G_{N_{F}}^{(4)}
(x_{1},x_{2},x_{3},x_{4}) 
 & = & \int d\mu^{(N_{F})}_{2}  
\bar{\lambda}^{\scriptscriptstyle\rm LD}_{\dot{\alpha}}(x_{1})
\bar{\lambda}^{\scriptscriptstyle\rm LD}_{\dot{\beta}}(x_{2})
\bar{\psi}^{\scriptscriptstyle\rm LD}_{\dot{\gamma}}
(x_{3})\bar{\psi}^{\scriptscriptstyle\rm LD}_
{\dot{\delta}}(x_{4})
\label{res}
\end{eqnarray}
The only remaining problem is to evaluate 
the $(28+4N_{F})$-dimensional\footnote{We are omitting the integral
over $x_{0}$ in this counting.} integral over the collective coordinates. 
The $\xi_{\alpha}$ and $\xi'_{\alpha}$ dependence in the four anti-fermion
fields (\ref{larged}) saturates the integrals over these parameters in
the measure. The remaining Grassmann integrals
must be saturated by expanding the exponential of the two-instanton
action or by the trilinear terms in (\ref{larged}). 
The bosonic integrals in (\ref{res}) 
can be performed using the same changes of variables used in the
$N_{F}=0$ case (see Sec.~8.2 of \cite{MO}). A straightforward 
calculation yields, 
\begin{eqnarray}
G_{N_{F}}^{(4)}(x_{1},x_{2},x_{3},x_{4})  & = & 
\frac{c_{N_{F}}}{g^{6}{\rm v}^{2}}\left(\frac{
\Lambda_{N_{F}}}{{\rm v}}\right)^{2(4-N_{F})}\int\, d^{4}x_{0}\,
\epsilon^{\alpha\beta}
{\rm S}_{\alpha\dot{\alpha}}(x_{1},x_{0})
{\rm S}_{\beta\dot{\beta}}(x_{2},x_{0}) \nonumber \\
& & \qquad{} \qquad{} \qquad{}  \, \times \, \epsilon^{\gamma\delta}
{\rm S}_{\gamma\dot{\gamma}}(x_{3},x_{0})
{\rm S}_{\delta\dot{\delta}}(x_{4},x_{0}) 
\label{final}
\end{eqnarray}
where $c_{0}=-945/8\pi^{2}$, $c_{1}=315/32\pi^2$,
$c_{2}=-15/64\pi^{2}$ and $c_{3}=c_{4}=0$. 
We can now compare this result with the contribution to $G^{(4)}_{N_{F}}$
from the pointlike four-fermion vertex in the exact low-energy effective
Lagrangian \cite{S,FP,MO}; 
\begin{eqnarray}
G_{N_{F}}^{(4)}(x_{1},x_{2},x_{3},x_{4})  & = & 
\frac{1}{8\pi i}\frac{\partial^{4}{\cal F}}
{\partial {\rm v}^{4}}
\int\, d^{4}x_{0}\,
\epsilon^{\alpha\beta}
{\rm S}_{\alpha\dot{\alpha}}(x_{1},x_{0})
{\rm S}_{\beta\dot{\beta}}(x_{2},x_{0}) \nonumber \\
& & \qquad{} \qquad{}  \, \times \, \epsilon^{\gamma\delta}
{\rm S}_{\gamma\dot{\gamma}}(x_{3},x_{0})
{\rm S}_{\delta\dot{\delta}}(x_{4},x_{0}) 
\label{4f}
\end{eqnarray}
Substituting the expansion (\ref{pre}) for ${\cal F}$ in (\ref{4f})
and comparing the two-instanton term with (\ref{final}) we extract 
the values ${\cal F}_{2}^{(0)}=5/2^{4}$, ${\cal
F}_{2}^{(1)}=-3/2^{5}$ and ${\cal F}_{2}^{(2)}=1/2^{6}$ in precise
agreement with the exact solution of Seiberg and Witten. 
\paragraph{}
In the exact solution, the two-instanton 
contribution to the four-point function vanishes automatically for 
$N_{F}=3$ and $4$. This follows from 
the action of the four derivatives with respect to
${\rm v}$ in (\ref{4f}) on the corresponding terms  
in the prepotential (\ref{pre}). 
In fact, the vanishing of the semiclassical result (\ref{final}) for
$N_{F}=3$ and $4$ has a similar
explanation. Because of the asymptotic forms (\ref{larged}) of the
anti-fermion fields, 
the semiclassical expression (\ref{res}) for $G^{(4)}_{N_{F}}$ also contains
four derivatives with respect to ${\rm v}$. 
By interchanging the order of these
derivatives and the collective coordinate integrations in (\ref{res}) 
it is easy to show that $c_{3}=c_{4}=0$ in (\ref{final}). 
\paragraph{} 
In the remainder of this paper, we will briefly 
indicate how these results can be
generalized to include non-zero masses, 
$m_{N_{F}}\geq m_{N_{F}-1}\geq\ldots\geq m_{1}$, for the hypermultiplets. 
We begin by describing the proposed exact results which we wish to
test. As in the massless case, 
the instanton expansion of the exact prepotential can be deduced
from the elliptic curves given in \cite{SW2} (see also \cite{OH}). 
An important restriction on the form of the prepotential
comes from the decoupling limit $m_{N_{F}}\rightarrow\infty$. In this
limit the exact prepotential obeys; 
\begin{equation} 
{\cal F}^{(N_{F})}({\rm v}; \{ m_{i} \},\Lambda_{N_{F}}) \rightarrow 
{\cal F}^{(N_{F}-1)}({\rm v}; \{ m_{i}\},\Lambda_{N_{F}-1})
\label{decouple}
\end{equation}
where the $\Lambda$-parameters for different numbers of flavours are
related as, 
\begin{eqnarray}
m_{N_{F}}\Lambda^{4-N_{F}}_{N_{F}} & \rightarrow
&\Lambda^{5-N_{F}}_{N_{F}-1}
\end{eqnarray}
with $\Lambda_{0}$ being the $\Lambda$-parameter of $N=2$ SYM theory. 
Note that the relation (\ref{decouple}) holds independently at each 
order in the instanton expansion. 
\paragraph{}
When all the masses are non-zero, the prepotential gets 
contributions from all numbers of instantons. 
In particular, there is a one-instanton contribution; 
\begin{equation}
\left. {\cal F}^{(N_{F})}({\rm v};
\{m_{i}\},\Lambda_{N_{F}})\right|_{n=1}=-\frac{i}{\pi}
\frac{\Lambda_{N_{F}}^{4-N_{F}}}{{\rm v}^{2}}{\cal F}^{(0)}_{1}\,
\prod_{j=1}^{N_{F}}m_{j}
\label{1I}
\end{equation}
This clearly obeys the decoupling relation (\ref{decouple}) 
and, in particular,  
the coefficient ${\cal F}^{(0)}_{1}$ can be extracted from the instanton
expansion of the prepotential of the $N_{F}=0$ theory: 
${\cal F}^{(0)}_{1}=1/2$ \cite{FP,MO}. 
\paragraph{}
The exact two-instanton contribution 
can be expanded in terms of the following set of 
$SO(2 N_{F})$-invariant polynomials in the masses, $m_{i}$; 
\begin{eqnarray}
M^{(N_{F})}_{0} &\qquad{} = \qquad{}  & 1     \nonumber \\
M^{(N_{F})}_{1} &\qquad{} = \qquad{}  & \sum_{i=1}^{N_{F}} m_{i}^{2}
\nonumber \\  
M^{(N_{F})}_{2} & =& \sum_{i<j}^{N_{F}} m^{2}_{i}m^{2}_{j}    \nonumber \\
\vdots    \qquad{}      &  & \vdots  \nonumber \\       
M^{(N_{F})}_{N_{F}} & = & \prod_{j=1}^{N_{F}} m_{j}^{2}    
\label{inv}
\end{eqnarray}
Explicitly we have, 
\begin{equation}
\left. {\cal F}^{(N_{F})}({\rm v};\left\{ m_{i} \right\},
\Lambda_{N_{F}})\right|_{n=2}
= -\frac{i}{\pi}{\rm v}^{2}\left(\frac{\Lambda_{N_{F}}}
{{\rm v}}\right)^{2(4-N_{F})}
\sum_{\delta =0}^{N_{F}} {f}^{(N_{F})}_{\delta}\left(\frac{
M_{\delta}^{(N_{F})}}{{\rm v}^{2\delta}}\right)
\label{exp}
\end{equation}
In the chiral limit, $m_{i}\rightarrow 0$, we
recover the coefficients of (\ref{pre}), hence we have 
$f^{(N_{F})}_{0}={\cal F}^{(N_{F})}_{2}$. 
The decoupling relation (\ref{decouple})
implies that the numerical coefficients
$f^{(N_{F})}_{\delta}$ are not independent but obey; 
\begin{equation}
f^{(N_{F})}_{\delta}=f^{(N_{F}-1)}_{\delta -1}=\ldots=
f^{(N_{F}-\delta)}_{0}
\label{r1}
\end{equation}
It follows that the constant $f^{(N_{F})}_{\delta}$ is equal to the 
coefficient ${\cal F}^{(N_{F}-\delta)}_{2}$ of the massless 
case. Hence, to extend our check of 
the two-instanton prediction to the massive case, 
it suffices to verify (\ref{exp}) and (\ref{r1}).  
\paragraph{}
The inclusion of a mass term in the instanton calculation is
straightforward. For any number of instantons, 
the only modification is to introduce an additional
dependence on the Grassmann parameters ${\cal K}_{li}$ and
$\tilde{\cal K}_{li}$ in the instanton action. 
Specifically, the mass term for the fundamental fermions
$\chi$ and $\tilde{\chi}$ can be evaluated on the zero modes
(\ref{fund}) using (\ref{orth}) and gives the additional term; 
\begin{equation}
S_{\rm mass}=\pi^{2}\sum_{i=1}^{N_{F}}m_{i}{\cal
K}_{li}\tilde{\cal K}_{li}
\label{mass}
\end{equation}
The expressions (\ref{1I}), (\ref{exp}) and (\ref{r1}) can be verified by 
calculating the corresponding one- and two-instanton 
contributions to $G^{(4)}_{N_{F}}$ as 
before. The only modification to the massless calculation 
comes from expanding the exponential $\exp(-S_{\rm mass})$ prior to 
performing the integrals over ${\cal K}_{li}$ and $\tilde{\cal
K}_{li}$. More complete details of all the calculations described above will
appear in the near future \cite{MOII}.    
\paragraph{}
ND is supported by a PPARC Advanced Research Fellowship, ND and VVK
are supported by the Nuffield foundation. MPM is supported by the
Department of Energy. 

\paragraph{}


\begin{thebibliography}{99}
\bibitem{SW1} N. Seiberg and E. Witten, 
{\it Electric-magnetic duality, monopole
condensation, and confinement in $N=2$ supersymmetric Yang-Mills theory}, 
{\it Nucl. Phys.} {\bf B426} (1994) 19, (E) {\bf B430} 
(1994) 485  hep-th/9407087. 
\bibitem{SW2} N. Seiberg and E. Witten, 
{\it Monopoles, duality and chiral symmetry breaking
in $N=2$ supersymmetric QCD}, 
{\it Nucl. Phys} {\bf B431} (1994) 484 ,  hep-th/9408099. 
\bibitem{FP} D. Finnell and P. Pouliot,
{\it Instanton calculations versus exact results in 4 dimensional 
SUSY gauge theories},
{\it Nucl. Phys.} {\bf B453} (1995) 225, hep-th/9503115. 
\bibitem{MO} N. Dorey, V. V. Khoze and M. P. Mattis, {\it
Multi-instanton calculus in $N=2$ supersymmetric gauge theory},
hep-th/9603136, to appear in {\it Phys. Rev.} {\bf D}. 
\bibitem{MOII} N. Dorey, V. V. Khoze and M. P. Mattis,
{\it Multi-instanton calculus in $N=2$ supersymmetric gauge theory II:
coupling to matter}, SWAT/125, preprint in preparation.
\bibitem{IT} K. Ito and S. K. Yang, {\it Picard-Fuchs equations and
prepotentials in $N=2$ supersymmetric QCD}, hep-th/9603073, to appear in
Proc. ``Frontiers of quantum field theory'', Toyonaka, Japan (1995).  
\bibitem{S}  N. Seiberg, {\em Phys. Lett.} {\bf  B206} (1988) 75.
\bibitem{ADS} I. Affleck, M. Dine and N. Seiberg, {\it Nucl. Phys.} {\bf B241}
(1984) 493; {\it Nucl. Phys.} {\bf B256} (1985) 557.  
\bibitem{AF} I. Affleck, {\it Nucl. Phys.} {\bf B191} (1981) 429. 
\bibitem{ADHM}  M. F. Atiyah, V. G. Drinfeld, N. J. Hitchin and
Yu. I. Manin,{\it Phys. Lett.} {\bf  A65} (1978) 185. 
\par V. G. Drinfeld and Yu. I. Manin, {\it Commun. Math. Phys.} 
{\bf 63} (1978) 177.
\bibitem{WB} J. Wess and J. Bagger, {\it Supersymmetry and supergravity}, 
Princeton University Press, 1992.
\bibitem{CWS} N. H. Christ, E. J. Weinberg and N. K. Stanton, {\it
Phys Rev} {\bf D18} (1978) 2013. 
\bibitem{COR1} E. Corrigan, P. Goddard and S. Templeton,
{\it Nucl. Phys.} {\bf B151} (1979) 93. 
\bibitem{COR2} E. Corrigan, D. Fairlie, P. Goddard and S. Templeton,
{\it Nucl. Phys.} {\bf B140} (1978) 31.
\bibitem{O1} H. Osborn, {\it Nucl. Phys.} {\bf B140} (1978) 45. 
\bibitem{O2} H. Osborn, {\it Ann. Phys.} {\bf 135} (1981) 373. 
\bibitem{OH} Y. Ohta {\it Prepotentials of $N=2$ $SU(2)$ Yang-Mills
gauge theory coupled with a massive matter multiplet}, hep-th/9604051. 
\par Y. Ohta {\it Prepotentials of $N=2$ $SU(2)$ Yang-Mills
theories coupled with massive matter multiplets}, hep-th/9604059.  
\end{thebibliography}
\end{document}